\documentclass[conference]{IEEEtran}
\IEEEoverridecommandlockouts
\usepackage{cite}
\usepackage{amsmath,amssymb,amsfonts}
\usepackage{algorithmic}
\usepackage{graphicx}
\usepackage{textcomp}
\usepackage{xcolor}
\usepackage{bibentry}
\def\BibTeX{{\rm B\kern-.05em{\sc i\kern-.025em b}\kern-.08em
    T\kern-.1667em\lower.7ex\hbox{E}\kern-.125emX}}
\usepackage{times}  
\usepackage{helvet}  
\usepackage{courier}  
\usepackage[hyphens]{url}  
\usepackage{graphicx} 
\usepackage{caption} 
\usepackage{algorithm}
\usepackage{algorithmic}
\usepackage{xcolor}

\usepackage{newfloat}
\usepackage{listings}

\usepackage{multicol}
\usepackage{multirow}
\usepackage{booktabs}

\DeclareCaptionStyle{ruled}{labelfont=normalfont,labelsep=colon,strut=off} 
\floatstyle{ruled}
\newfloat{listing}{tb}{lst}{}
\floatname{listing}{Listing}
\begin{document}

\title{Automated Trading System for Straddle-Option Based on Deep Q-Learning
}

\author{\IEEEauthorblockN{1\textsuperscript{st} Yiran Wan}
\IEEEauthorblockA{\textit{Nankai University} \\
\textit{College of Software}\\
Tianjin, China \\
yiran.wan@mail.nankai.edu.cn}
\and
\IEEEauthorblockN{1\textsuperscript{st} Xinyu Ying}
\IEEEauthorblockA{\textit{Nankai University } \\
\textit{School of Finance }\\
Tianjin, China \\
xinyu.ying@mail.nankai.edu.cn}
\and
\IEEEauthorblockN{3\textsuperscript{rd} Shengze Xu}
\IEEEauthorblockA{\textit{The Chinese University of Hong Kong} \\
\textit{Department of Mathematics}\\
Hong Kong \\
szxu@math.cuhk.edu.hk}
}

\maketitle

\begin{abstract}
Straddle Option is a financial trading tool that explores volatility premiums in high-volatility markets without predicting price direction. Although deep reinforcement learning has emerged as a powerful approach to trading automation in financial markets, existing work mostly focused on predicting price trends and making trading decisions by combining multi-dimensional datasets like blogs and videos, which led to high computational costs and unstable performance in high-volatility markets. To tackle this challenge, we develop automated straddle option trading based on reinforcement learning and attention mechanisms to handle unpredictability in high-volatility markets. Firstly, we leverage the attention mechanisms in Transformer-DDQN through both self-attention with time series data and channel attention with multi-cycle information. Secondly, a novel reward function considering excess earnings is designed to focus on long-term profits and neglect short-term losses over a stop line. Thirdly, we identify the resistance levels to provide reference information when great uncertainty in price movements occurs with intensified battle between the buyers and sellers. Through extensive experiments on the Chinese stock, Brent crude oil, and Bitcoin markets, our attention-based Transformer-DDQN model exhibits the lowest maximum drawdown across all markets, and outperforms other models by 92.5\% in terms of the average return excluding the crude oil market due to relatively low fluctuation.
\end{abstract}


\section{Introduction}

The high-volatility market, which involves over \$40 trillion in market capitalization as well as significantly higher risk and profit opportunity, has attracted the attention of innumerable investors around the world. The straddle option, which is designed for scenarios where a trader anticipates significant market volatility but is uncertain about the direction of the price movement, naturally applies to the high-volatility markets \cite{brenner2006hedging}. Meanwhile, Deep Reinforcement Learning (DRL) has achieved significant success in various quantitative trading tasks, particularly in trades based on predicting price direction \cite{FAN2022}. However, the characteristics of high volatility markets, including market uncertainty, model adaptability issues, sensitivity to future events, and the inability to capture all factors affecting stock prices, limit the accuracy of these predictions \cite{gondkar2021stock,Boukherouaa2021,choudhury2014real,vui2013review} and often result in extreme losses during 'black swan' events. Conversely, straddle options that focus on trading volatility can mitigate potential losses caused by inaccurate stock price predictions. In fact, existing works have already demonstrated the distinguished performance of straddle option in high-volatility markets and comprehensively discussed the performance of long and short straddle strategies under different market conditions, highlighting that long straddle options can effectively hedge risks and offer potential high returns in high volatility markets \cite{shivaprasad2022choosing}. Even though, few papers have leveraged straddle options in algorithmic trading for high-volatility markets. Therefore, we introduce the notion of straddle option in algorithmic trading to pursue stable excess return in high volatility markets in the long run.
To hunt optimal timing for the straddle option to open and close the position, we face the following two major challenges: i) How to hunt the optimal timing to trade and adopt different strategies at various points of market volatility; ii) How to enable the model to understand long-term trends while focusing on short-term fluctuations to achieve long-term gains. To tackle these two challenges, we apply Transformer-DDQN model and design a new attention network method to calculate the Q-value for trading decisions. Our method allows for dynamic adjustments to different market conditions and long-term excess return maintaining with a more stable and superior performance. Our contributions are four-fold:
\begin{itemize}
    \item We design the attention mechanism in two approaches and employ DDQN framework for the optimization of trading strategies. \\
    i) Self-attention mechanism ensures rapid acquisition of the latest market information and optimizes asset weights according to the current market information;\\
    ii) Channel attention mechanism allows the model to balance short-term reactive adjustments with a strategic understanding of long-term trends, thereby optimizing for long-term gains.
    \item We input the reference information of resistance level, which indicates potential points of price reversal or consolidation, to make more informed decisions about when to enter or exit trades.
    \item We opt for a delayed reward function with a stop line to avoid the problem of getting stuck in local optima or excess loss.
    \item Experiments on the Chinese stock, Brent crude oil, and Bitcoin markets demonstrate the superiority of the attention-based Trans-DDQN model for straddle option trading over six baseline models in terms of three widely recognized financial metrics.
\end{itemize}

\section{Relate Work}
\subsection{Deep Reinforcement Learning}
Deep Reinforcement Learning has emerged as a powerful approach for automating trading strategies in financial markets. AbdelKawy et al. \cite{abdelkawy2021synchronous} proposed a multi-stock trading model using a synchronous multi-agent DRL approach. This model dynamically extracts financial data features and applies scalable DRL techniques to handle large historical trading datasets. Tran et al. \cite{tran2022parameter} optimized parameters for trading strategies using DRL, specifically Double Deep Q-Network (DDQN) and Bayesian Optimization. Their system, applied to cryptocurrency markets, achieved positive returns and outperformed other optimization methods in daily trading scenarios. Azhikodan et al. \cite{azhikodan2019stock} developed a swing trading bot using a Deep Deterministic Policy Gradient (DDPG) model. They combined DRL with sentiment analysis from financial news to predict stock trends and make trading decisions. Kabbani et al. \cite{kabbani2022deep} applied the Twin Delayed Deep Deterministic Policy Gradient algorithm to automate trading. Their model, formulated as a Partially Observed Markov Decision Process, achieved a high Sharpe ratio, outperforming traditional machine learning approaches.
\subsection{Attention Networks}
Attention mechanisms have significantly enhanced the performance, efficiency, and interpretability of neural networks by enabling models to selectively focus on relevant input features across a variety of tasks, from image localization and understanding to sequence-based models \cite{itti2001computational,olshausen1993neurobiological,jaderberg2015spatial,bluche2016joint}. The Transformer model leverages self-attention mechanisms instead of recurrent or convolutional layers, enabling parallelization and more effective handling of long-range dependencies in sequences \cite{vaswani2017attention}. The SAGAN model by \cite{zhang2019self} incorporates self-attention mechanisms into the GAN framework, enhancing the generation of high-resolution images.  Hu, Shen, and Sun \cite{ChannelAttention} presented "Squeeze-and-Excitation Networks" (SE-Nets), which increase the representational power of networks by adaptively recalibrating channel feature responses. 
Building on these advancements, our model uses self-attention mechanisms for time series data to quickly capture the latest market information and optimize asset weights accordingly. We also integrate channel attention mechanisms to balance short-term adjustments with a strategic view of long-term trends for higher long-term gains. This combined approach aims to enhance trading performance by leveraging the strengths of both self-attention and channel attention.

\section{Problem formalization}

\subsection{Simulation of an options trading environment}
\paragraph{Historical volatility}
Standard variance is a common measure of asset price volatility, and its formula is \cite{OptionBook}:
\begin{equation}
    \sigma=\sqrt{\frac{\sum_{i=1}^N(r_i-\bar r)^2}{N-1}}
    \label{Eq.1}
\end{equation}
where $r_i$ represents the logarithmic return for the period, $\bar r$ is the average return of the sample sequence, and $N$ is the sample size. To standardize this expression to annualized volatility, we need to multiply it by the annualization factor $\sqrt F$, where $F$ is the number of periods in a year.

In real financial markets, it is quite challenging to separate the mean and volatility components of returns, especially in the case of small samples, the estimation of the mean of return is very inaccurate \cite{Option-Volatility}. This paper assumes that the probability of the market rising or falling within a short period is 50\%, thus the average market return over this period is zero. For estimating historical volatility, we use the closing prices of each candlestick data, resulting in the final Equation \ref{Eq.2}:

\begin{equation}
    HV=\sqrt{\frac{F}{N-1}\sum_{i=1}^N[ln(\frac{c_i}{c_{i-1}})]^2}
    \label{Eq.2}
\end{equation}
where $c_i$ is the closing price at time $i$.

The calculation of volatility using standard deviation is straightforward and easy to understand, making it one of the most common methods for measuring volatility. This paper utilizes 15-minute candlestick data, which allows capturing intraday volatility even in sideways trading markets.

\paragraph{Profit and Loss Settlement Method}
The price of an option is primarily influenced by three factors: the current price $S$, the remaining time to expiration $T$ (annualized), and the volatility $\sigma$ . The risk-free interest rate is generally considered constant in the short term. The Black-Scholes model, developed by American financial economists Fisher Black and Myron Scholes in 1973, is an option pricing model. By deriving and using the principle of European option parity, the pricing formula for European options without dividend payments can be obtained \cite{Option-BSM}:
\begin{equation}
    C(S,t)=SN(d_1)-Ke^{-r(T-t)}N(d_2) \label{Eq.3}
\end{equation}
\begin{equation}
    P(S,t)=-SN(-d_1)+Ke^{-r(T-t)}N(-d_2)\label{Eq.4} 
\end{equation}
where $d_1$ and $d_2$ are :
\begin{equation}
    d_1=\frac{ln(S/K)+(r+\sigma^2/2)(T-t)}{\sigma \sqrt{T-t}} \label{Eq.5}
\end{equation}
\begin{equation}
    d_2=d_1-\sigma\sqrt{T-t}\label{Eq.6} 
\end{equation}

Here, $C(S,t)$ represents the price of a European call option, and $P(S,t)$ represents the price of a European put option. $T-t$ denotes the annualized time to expiration (generally defined as the remaining days divided by 360 days). $r$ represents the risk-free interest rate, $K$ is the strike price, and $N(d)$ is the cumulative distribution function of the standard normal distribution, i.e., the probability that the variable is less than $d$. The Black-Scholes pricing model is widely accepted in the options market and serves as an important reference for many investors.

This paper primarily investigates short-term trading of monthly options. Since the exchange adjusts the strike price and contract multiplier of an ETF in the event of dividends, the impact of dividends on option pricing is ignored. Implied volatility typically fluctuates around historical volatility. Due to the difficulty in obtaining high-frequency data and its susceptibility to market sentiment, this paper assumes a risk-neutral market and uses historical volatility from the past $n$ days as a substitute for implied volatility in profit and loss settlement.

\paragraph{Rules for Establishing Straddle Option Positions}
Set the interval of the strike price as $S$, so the price $P$ will necessarily fall within the interval $[X,X+S]$, where $X=P-(P\,mod\,S)$. This interval is evenly divided into three smaller segments: $S1$, $S2$, and $S3$.

1. If the price falls within $S1$, choose the call and put options with a strike price of $X$.

2. If the price falls within $S2$, choose the call option with a strike price of $X+S$ and the put option with a strike price of $X$.

3. If the price falls within $S3$, choose the call and put options with a strike price of $X+S$. 

Once the contracts are selected, a certain number of both call and put options are purchased to construct a straddle option group, ensuring that the overall $delta$ of the portfolio is approximately zero.
Regarding the contract expiration date, this paper selects the options expiring in the current month. If the remaining time for the current month's options is less than 15 days, the next month's contracts are chosen to avoid the end-of-life effect of the options. Generally, options that are near-term and have a strike price close to the spot price have better liquidity and lower trading frequency, thus the slippage issue is temporarily ignored in this paper.

\subsection{Resistance Level Identification}
A resistance level is a crucial concept in technical analysis of the stock market \cite{Technical-Analysis1,Technical-Analysis2}. When the price reaches a resistance level, the battle between buyers and sellers intensifies, leading to greater uncertainty in price movements. In terms of trading psychology, people exhibit an anchoring effect, paying more attention to relative highs and lows. Here, multiple closely grouped swing highs/lows are defined as a resistance level. Identifying recent historical resistance levels on the candlestick chart can provide more reference information for the model. See Algorithmic\ref{AlgorithmicRes}, which automates the identification of resistance and support levels using historical candlestick data. By applying a sliding window and predefined thresholds, it consistently detects key price zones, providing reliable reference points for trading models and aligning with observed anchoring effects in trading psychology.

\begin{figure}[h]
\centering
\includegraphics[width=0.8\columnwidth]{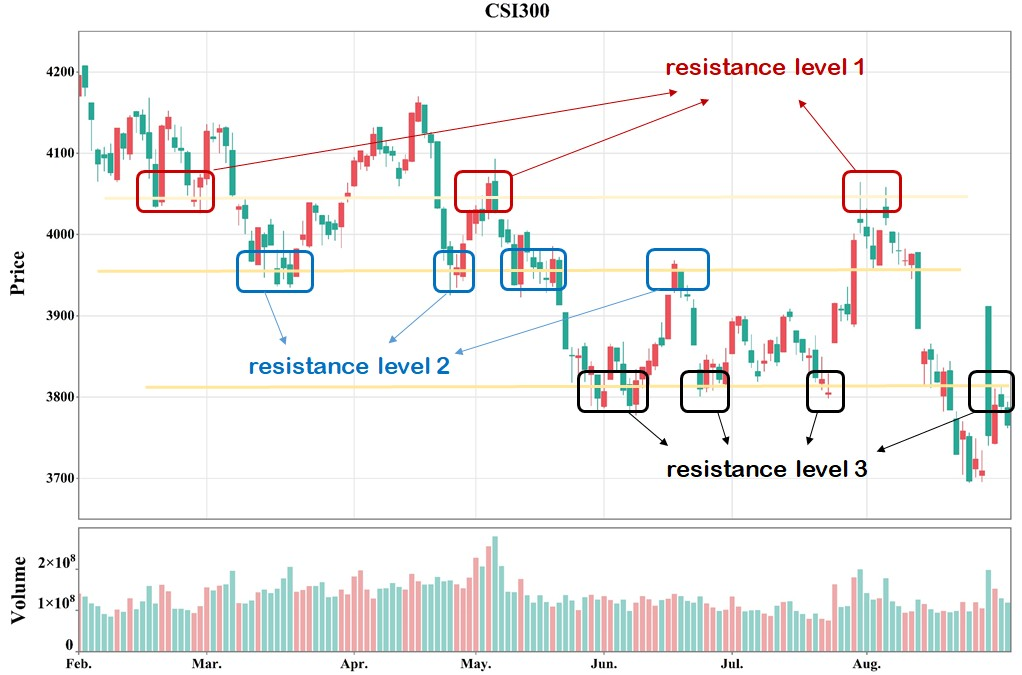}
\caption{This chart shows the historical market of the CSI 300 index from February to August 2023.}
\label{Resistance-Level}
\end{figure}

After obtaining the resistance levels, the area within ±0.3\% of these resistance levels is defined as the resistance area. When the price moves into the resistance area, a resistance signal is sent to the model, marked as 1; otherwise, it is marked as 0, as illustrated in Figure \ref{Resistance-Level}.
\begin{algorithm}
\begin{algorithmic}[1]
\caption{Get resistance and support points}
\label{AlgorithmicRes}
\REQUIRE Candlestick data (for the $i_{th}$ bar note as $bar_i$), length of sliding window $d$, reversal range $f\%$, and breakthrough range $e\%$;
\ENSURE resPointSet, supPointSet;
\STATE resPointSet=[($bar_0.close, bar_0.time$)];
\STATE supPointSet=[($bar_0.close, bar_0.time$)];
\FOR{$t=d$ to len(K-Bar DataSet)}
    \STATE $M_t=bar_t.close-bar_{t-d}.close$;
    \IF{$M_t*M_{t-1}<0$} 
        \IF{$M_t>0$}
           \STATE possibleSupportPoint=$bar_t$
            \IF{$\frac{bar_t.close}{resPointSet[-1,close]}-1<f\%$ \\
            \textbf{or}
             $\frac{bar_t.close}{supPointSet[-1,close]}-1<e\%$}
                \STATE supPointSet.add(possibleSupportPoint)
            \ENDIF
        \ENDIF
        \IF{$M_t<0$}
            \STATE possibleResistancePoint=$bar_t$
            \IF{$\frac{bar_t.close}{supPointSet[-1,close]}-1>f\%$ \\
            \textbf{or}
            $\frac{bar_t.close}{resPointSet[-1,close]}-1>e\%$}
               \STATE resPointSet.add(possibleResistancePoint)
            \ENDIF
        \ENDIF
    \ENDIF
\ENDFOR\\
\STATE \textbf{return} resPointSet, supPointSet
\end{algorithmic}
\end{algorithm}
\vspace{-5mm}
\vspace{-3mm}

\subsection{Definition of state and action space}
\label{SectionSt}
\paragraph{State definition}
The state features of the market at any given time are the core input of the model. A sliding window of length $d$ is set, taking the data from time $t-d$  to $t$ as the time series $Seq_t$. $Seq_t$ includes $d$ time steps $k_i$:
$Seq_t=[k_{t-d+1},k_{t-n+2},...k_t]^T$. Each time step  $k_i$ in $Seq_t$ includes the following information:
\begin{itemize}
    \item  Candlestick data enables the model to observe the information of the game between buyers and sellers in the market in the past period of time.
    \item The floating profit and loss of a single trade position, because it's essential for an investor to constantly monitor account information.
    \item Historical volatility over the recent $n$ days, as volatility is a significant factor in option pricing.
    \item The number of days until the next trading day, since the market does not trade on weekends and holidays, but the time value of options is still lost.
\end{itemize}

In summary, $Seq_t \in R^{f\times d}$. Additionally, there are two pieces of information given separately and not included in the time series data:
\begin{itemize}
    \item If the market price reaches a resistance level, the resistance level identification program will issue a resistance signal, noted as $ResistanceFlag$.
    \item The holding time of the position. It reminds the model of the time decay of holding options, noted as $HoldTime$.
\end{itemize}
All these components together form a state $S_t^1$:
\begin{equation}
    S_t^1=[Seq_t,ResFlag,HoldTime]
\end{equation}
where $S_t^1$ focuses on providing information for trading decisions.

We define the observation time series for market trends as $obs_t^p$, where $\bar p$ represents the period of the candlestick data, which can be 15 minutes, 30 minutes, 60 minutes, etc. The set of different period is denoted by $P$, and $t$ represents the current moment. Sliding window of length $d$ is used to extract data from different periods of candlestick data, allowing the model to observe market changes from various time scales. This component is referred to as $S_t^2$, focusing more on providing medium- to long-term market context information. Together, these two parts constitute the market state $S_t$

\paragraph{Action definition}

The action is the output result of the model. The action at time $t$ is $a_t\in A_t$. The action space has only two states: 1 representing the holding position and 0 representing the non-holding position. The transition from 0 to 1 signifies an opening action, and that from 1 to 0 signifies a closing action.
The model must execute the whole process, namely waiting without a position, opening a position, holding the position, and closing the position for a complete trade.
\begin{figure*}[h]
    \centering
    \includegraphics[width=0.95\textwidth]{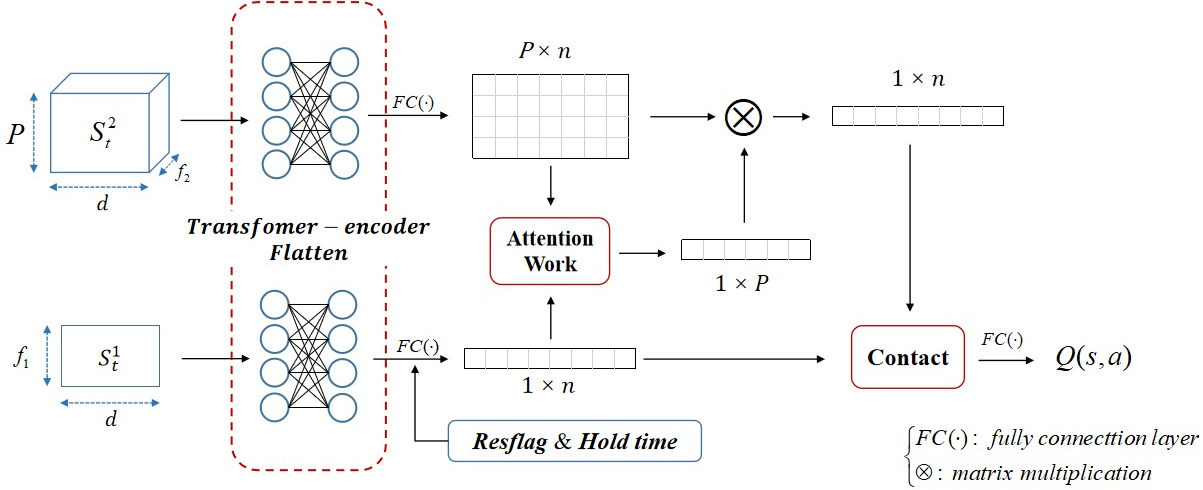}
    \caption{The network structure of the estimated Q-values. Firstly, the self-attention module is used to learn relevant information from the time series data. Then, the channel attention module is employed to integrate multi-period sequence information. Finally, the $Q(s_t,a_t)$ are output through fully connected layers.}
    \label{All-Structure}
\end{figure*}

\section{Method}
In the previous section, the market's state information is bifurcated into two segments labeled by $S_t^1$ and $S_t^2$, which are characterized by their temporal sequence nature and inter-sequence relationships. The neural network designed for Q-value estimation is accordingly compartmentalized into two distinct modules: one for processing time series information and another for integrating multi-period data.
\subsection{Time series information processing module}\label{TimeProcess}
Transformer is a deep learning model for natural language processing and sequence learning tasks, first introduced by \cite{normal-Transformer}. Its self-attention mechanism allows it to capture long-range dependencies in long sequences, which is crucial for sequence data tasks. Given that candlesticks data is also a type of time series, a Transformer-Encoder module can be used with a sliding window to learn the information contained in candlestick charts data.

Initially, the temporal sequence data $Seq_t$ is directly input into the Transformer-Encoder module to learn the market state information and to extract the corresponding features.
\begin{equation}
    H_t^1=Encoder(Seq_t,W)\,\,\, H_t^1\in R^{f\times d}\label{equ1.1}
\end{equation}

Afterward, the matrix is transformed into a vector by a flatten layer, and a dense layer is then applied to compress the information.
\begin{equation}
    H_t^2=\sigma(Flatten(H_t^1)W^{fd\times n})(fd>n) \,\,\,H_t^2 \in R^n \label{equ1.2}
\end{equation}

Then, we combine the $ResFlag$ and $HoldTime$:
\begin{equation}
    H_t^3=concat[H_t^2,ResFlag,HoldTime] \,\,\,H_t^3 \in R^{n+2}\label{equ1.3}
\end{equation}

Finally, a fully connected layer is employed to learn the $ResFlag$ and $HoldTime$ information.
\begin{equation}
    H_t^4=\sigma(H_t^3W_{(n+2)*n})\,\,\,H_t^4 \in R^{n}
\end{equation}

In this context, $f$ is the dimensionality of a single time step within the time series data, $d$ refers to the duration of the time window, and $n$ is the dimension of the embedding. For the observed market trend time series $obs_t^p$, the processing follows the same procedure as in Equations \ref{equ1.1}-\ref{equ1.2}, using a Transformer-encoder to learn the characteristics of the time series data denoted as $O_t^p$.

\subsection{Multi-period information fusion module}
In financial markets, short-term market trends can be similar, but the information they convey may differ in the context of different medium- and long-term trends. Therefore, trading should not solely focus on short-term candlestick chart information; it is also essential to consider the background information of long-term candlestick charts. Therefore, channel attention \cite{ChannelAttention} mechanism is employed to enhance the model's ability to process data across various cycles.

Regarding the multi-period information fusion module, $H^4_t$ is treated as the query vector 
$q$. The candlestick data from other periods, after being processed as described in \textit{Time series information processing module} and denoted as $O_t^p$, are used as both the key and value vectors $k$ and $v$, with $k=v$. Attention scores for each period relative to $H_t^4$ are computed and then normalized.
\begin{equation}
    a_p=\alpha(H_t^4,O_t^p) p \in P
    \label{equ2.1}
\end{equation}
\begin{equation}
    e_p=\frac{exp(a_p)}{\sum_{p \in P}exp(a_p)}\label{equ2.2}
\end{equation}
After obtaining the attention score, the information of multiple periods is fused.
\begin{equation}
    O_t=\sum_{p \in P}e_p\cdot O_t^p
\end{equation}
Finally, we have the integrated query vector output $Q(S_t,a_t)$,
\begin{equation}
    Q(S_t,a_t)=Liner(concat(H^4_t,O_t))
\end{equation}
When calculating attention scores, the operator $\alpha(\cdot)$ is $a=p^TWq$ \cite{BilinearAttention}. The overall structure of the network is shown in Figure \ref{All-Structure}.

\subsection{Design of the reward function}
The design of the reward function is one of the key factors affecting the performance of DRL-based models. Common reward functions include profit maximization, loss minimization, and risk-adjusted return maximization. Under different market volatilities, the reward function needs to be optimized to adapt to changing market conditions, such as by adjusting risk preference parameters to balance returns and risks. Millea \cite{millea2021deep} pointed out that using risk measures (e.g., Sharpe ratio and maximum drawdown) to design reward functions promotes a balance between risk control and return maximization. However, this direct performance assessment-based reward function is not suitable for straddle option trading, as frequent reward and penalty signals due to market fluctuations during the holding period can seriously interfere with model training and make it unstable. This paper uses a delayed reward mechanism to train the model and sets a stop-loss system to control drawdown risk.

Define the market value at the position at time $t$ as $MarketValue_t$, and the opening cost as $Cost$. The logarithmic return of this trade at time $t$ is $return_t=ln(\frac{MarketValue_t}{Cost})$ and the stop-loss threshold is set as $stop$ (where $stop<0$).
\begin{itemize}
    \item If $a_{t-1}\rightarrow a_t$ is $0\rightarrow1$.  The movement is defined as opening a position and $reward_t=0$.
    \item If $a_{t-1}\rightarrow a_t$ is $1\rightarrow1$. The movement is defined as holding a position, which leads to 2 scenarios:\\
    1. If $return_t>stop$, then $reward_t=0$.\\
    2. If $return_t<stop$, then $reward_t=e^{return_t}-1$ (converted to a simple return).\\
    This design allows for a certain amount of loss, enabling the model to ignore short-term noise and focus more on the market fluctuations over a period of time.
    \item if $a_{t-1}\rightarrow a_t$ is $1\rightarrow0$. The movement is defined as closing a position, which leads to 2 scenarios:\\
    1. If closing occurs at the stop-loss threshold, $reward_t=a$ (where $a>0$), as stopping the loss is the correct action.\\
    2. Otherwise, $reward_t=e^{return_t}-1$, and if the closing point during profit-taking deviates by more than $g$\% from the opening point, a double reward is given to encourage the model to hold the position firmly in extreme unidirectional market conditions.
    \item If $a_{t-1}\rightarrow a_t$ is $0\rightarrow0$. The movement is defined as not holding a position and $reward_t=0$.  
\end{itemize}
By using this reward function, the model learns to manage straddle option positions effectively, maintaining stability while responding to market fluctuations. The delay-reward and stop-loss mechanisms benefit in filtering out noise and focusing on significant market movements.

\begin{table*}[h]
\setlength{\tabcolsep}{1mm}
  \centering
  \begin{tabular}{cccccccccccccccccccc}
    \toprule
    \multirow{2}{*}{Method}&\multicolumn{3}{c}{SSE50}&&\multicolumn{3}{c}{CSI300}&&\multicolumn{3}{c}{CSI500}&&\multicolumn{3}{c}{Brent Crude}&&\multicolumn{3}{c}{BTC}\\
    \cmidrule(r){2-4}\cmidrule(r){6-8}\cmidrule(r){10-12}\cmidrule(r){14-16}\cmidrule(r){18-20}
     &MR&SP&MDD && MR&SP&MDD && MR&SP&MDD&& MR&SP&MDD&& MR&SP&MDD\\
    \cmidrule(r){1-20}
    Long&-0.11&-0.71&-0.38&&-0.12&-0.74&-0.39&&-0.11&0.68&-0.47&&\textbf{0.05}&\textbf{0.03}&-0.64&&0.19&0.17&-1.13\\
    MA&-0.01&-0.33&-0.29&&0.01&-0.31&-0.28&&0.02&-0.22&-0.28&&-0.29&-0.68&-1.22&&0.30&0.39&-0.66\\
    \cmidrule(r){1-20}
    Xgboost&-0.33&-1.82&-0.76&&-0.28&-1.65&-0.68&&-0.15&-0.93&-0.42&&-0.44&-0.97&-1.35&&-0.14&-0.31&-1.18\\
    LSTM&-0.04&-0.43&-0.28&&-0.19&-1.02&-0.43&&-0.21&-1.02&-0.58&&-0.04&-0.1&-1.13&&-0.29&-0.66&-1.31\\
    \cmidrule(r){1-20}
    GRU-DDQN&-0.46&-2.12&-1.04&&-0.38&-1.83&-0.97&&-0.59&-2.49&-1.45&&-1.03&-0.19&-2.36&&-0.96&-0.14&-2.38\\
    DDPG&-1.31&-6.03&-2.99&&-1.60&-7.89&-3.59&&-1.27&-5.42&-2.88&&-1.05&-0.12&-2.41&&-1.04&-0.12&-2.31\\
    \textbf{Trans-DDQN}&\textbf{0.45}&\textbf{1.03}&\textbf{-0.15}&&\textbf{0.42}&\textbf{1.75}&\textbf{-0.15}&&\textbf{0.60}&\textbf{1.22}&\textbf{-0.25}&&-0.07&-0.33&\textbf{-0.53}&&\textbf{0.72}&\textbf{2.27}&\textbf{-0.11}\\    
    \bottomrule
  \end{tabular}
\caption{Performance comparison with baseline on data set.}
\label{BaselineExperimentsTable}
\end{table*}

\begin{figure*}[h]
\begin{center}
\begin{minipage}[b]{0.195\linewidth}
\includegraphics[width=\textwidth]{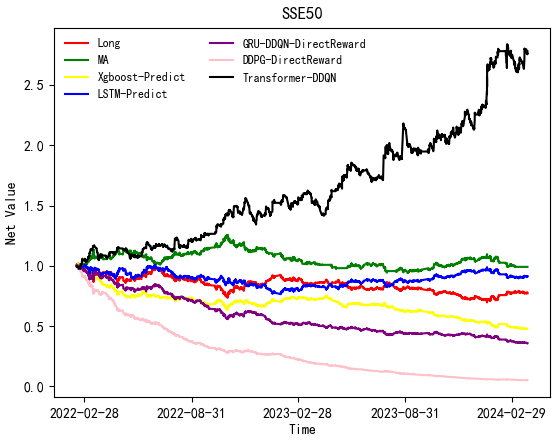}
\end{minipage}
\hfill
\begin{minipage}[b]{0.195\linewidth}
\includegraphics[width=\textwidth]{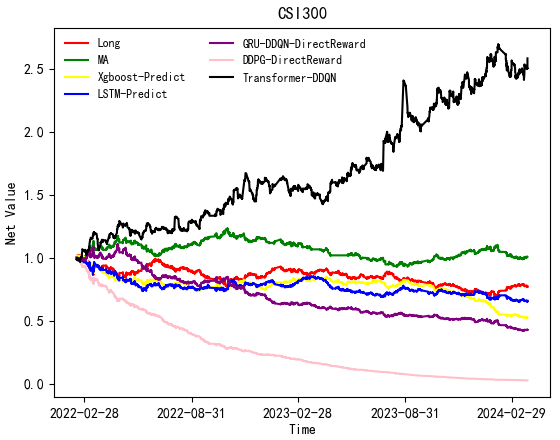}
\end{minipage}
\hfill
\begin{minipage}[b]{0.195\linewidth}
\includegraphics[width=\textwidth]{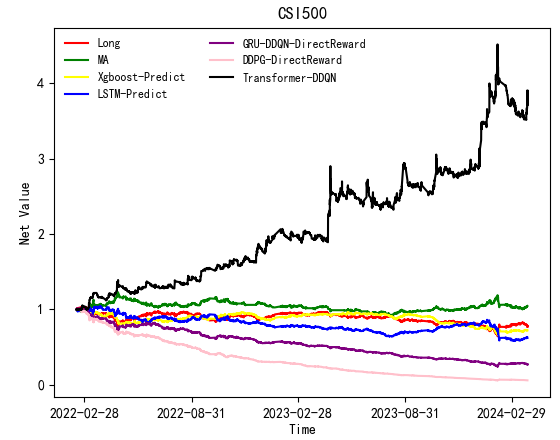}
\end{minipage}
\hfill
\begin{minipage}[b]{0.195\linewidth}
\includegraphics[width=\textwidth]{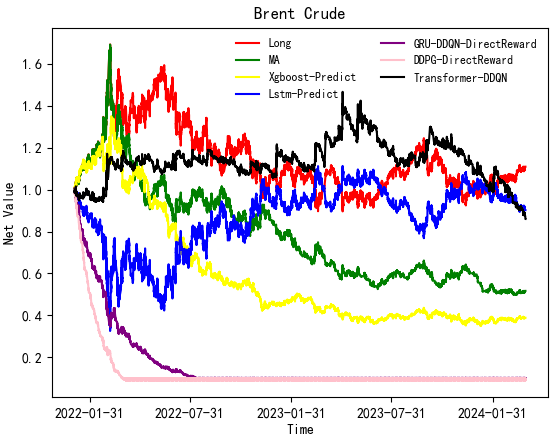}
\end{minipage}
\hfill
\begin{minipage}[b]{0.195\linewidth}
\includegraphics[width=\textwidth]{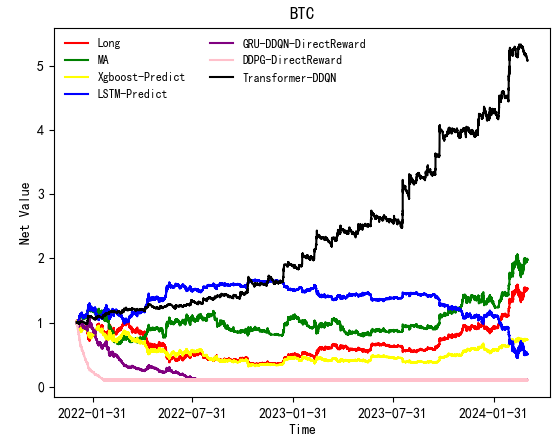}
\end{minipage}

\caption{Performance comparison with 6 baselines}
\label{BselineResult}
\end{center}
\vspace{-5mm}
\end{figure*}

\section{Experiment}
\subsection{Dateset description}
The data used in this paper are the main broad-based indices traded on the Shanghai Stock Exchange(SSE), specifically the SSE 50, CSI 300, and CSI 500. To test the generalizability of the method, experiments are also conducted on Brent crude and Bitcoin datasets. We collect 15-minute candlestick data for these assets from January 4, 2018, to March 31, 2024. Each data point includes the open-high-low-close price, volume, and trading value for the period. In the A-share market, index option products tracked include ETF options listed on the SSE and index options traded on the China Financial Futures Exchange (CFFEX). For convenience, they are collectively referred to as index options in this study.
\subsection{Experimental Environment Setup}
The 15-minute candlestick data from January 1, 2018, to December 31, 2021, is designated as the training set, while the data from January 1, 2022, to March 31, 2024, is used as the testing set. The model is configured to look back at historical data over a period of 20 days, and historical volatility is calculated based on the past 5 days. To better simulate real market conditions, the following constraints are added: brokerage fees for index options, as charged by the CFFEX, are 15 RMB per contract; however, since the contract size for CFFEX index options is 100, the corresponding fee per point is 0.15 RMB. To mitigate risk and prevent excessive speculation, the exchange imposes a position limit, which restricts the contract value to no more than 20\% of total funds when opening a new position. On Binance, the trading fee for bitcoin options is 0.02\% of the strike price times the number of contracts size, but it cannot exceed 10\% of the option premium. For Brent crude oil options traded on the London ICE, the trading fee is 1.5 USD per contract, with a contract size of 1,000. The initial capital for the experiment is set at 1 million. In the baseline experiment, the trading cost for ETFs is set at 0.05\% of the total transaction value. Since the focus is on short-term trading, the maximum holding period for options is limited to 5 days to avoid high time decay costs. In Section \textit{Design of the reward function}, the stop-loss threshold is set at 15\%.

The chosen evaluation metrics include: Annualized average logarithmic return (AVGR), Sharpe ratio (SP), and Maximum drawdown (MDD) in logarithmic form.

\subsection{Experiment design}
\paragraph{Baseline methods}
The study compares the proposed model against 2 rule-based trading strategies: Market's Own Return and Dual Moving Average Strategy \cite{gunasekarage2001MA}, 2 stock price prediction models based on machine learning: XGBoost \cite{chen2016xgboost} and LSTM Network \cite{LSTM1-DLPredict} for daily stock price prediction, and 2 deep reinforcement learning-based automated trading models: GRU-DDQN \cite{GRU-DDQN1-RL} and DDPG \cite{normal-DDPG}. 

\paragraph{Ablation study design}
An ablation study is conducted on the test data to show how different components of the model affect the final results. Three model variants are selected, each altering one component of the model:
\begin{itemize}
    \item NoRes-Transformer-DDQN: The resistance level information is masked, but all other components remain unchanged.
    \item DR-Transformer-DDQN: Use common performance metrics (returns) as the reward function, with all other components unchanged.
    \item LSTM-DDQN: Replace the Transformer-Encoder part used to learn $Seq_t$ with an LSTM network for estimating $Q(s_t,a_t)$, with all other components kept unchanged.
\end{itemize}

\begin{table*}[h]
  \centering
  \begin{tabular}{cccccccccccc}

    \toprule
    \multirow{2}{*}{Method}&\multicolumn{3}{c}{SSE50}&&\multicolumn{3}{c}{CSI300}&&\multicolumn{3}{c}{CSI500}\\
    \cmidrule(r){2-4}\cmidrule(r){6-8}\cmidrule(r){10-12}
     &MR&SP&MDD && MR&SP&MDD && MR&SP&MDD\\
    \cmidrule(r){1-12}
    NoRes&0.2245&0.4106&-0.1595&&0.3852&1.7134&-0.1925&&0.3458&1.1075&-0.1300\\
    DR&0.2162&0.5680&-0.1599&&0.1417&0.7578&-0.1498&&0.2440&0.8033&-0.1274\\
    LSTM&0.4106&0.8148&-0.2078&&0.3884&1.7066&-0.1818&&0.5572&\textbf{1.6324}&\textbf{-0.1016}\\
    \textbf{Trans-DDQN}&\textbf{0.4542}&\textbf{1.0328}&\textbf{-0.1518}&&\textbf{0.4214}&\textbf{1.7460}&\textbf{-0.1475}&&\textbf{0.6049}&1.2192&-0.2509\\  
    \bottomrule
  \end{tabular}
\caption{Ablation experiments result}
\label{AblationExperimentsTable}
\vspace{-3mm}
\end{table*}

\subsection{Experimental results and analysis}

\paragraph{Comparison with baseline results}
The performance of the proposed model is compared with the baseline methods on the data set. The net performance is depicted in Figure \ref{BselineResult} and the evaluation index is shown in Table \ref{BaselineExperimentsTable}.

It is found that the Transformer-DDQN proposed in this paper outperforms the baseline methods based on transaction price direction on various performance indicators. Due to the high volatility of the A-share market, once the trading direction is wrong, large losses are encountered. Therefore, Transformer-DDQN turns to focus on the information about price fluctuations, in order to find potential profit opportunities in a highly uncertain market. The rule-based trading strategies are only suitable for some trading conditions, and in particular, the double-moving average (MA) strategy performs well in trending conditions, but is prone to losses in volatile conditions. The reason that Xgboost-Predict and LSTM-Predict trade poorly based on stock price prediction is that they only pursue the accuracy of prediction but ignore the trading odds factor. It is impossible to completely avoid losses in trading, and profits must be the result of a combination of winning rates and odds. GRU-DDQN with direct reward and DDPG show that the training has failed. Due to the high volatility of the market, profit and loss frequently alternate, resulting in frequent switching of rewards and penalties during the training and increased difficulty for the model to learn valuable information.

It is also found that the performance of Transformer-DDQN varies across different datasets. In the A-share market, its profitability for the SSE 50 and CSI 300 is weaker compared to that for the CSI 500. This is because the primary components of the SSE 50 and CSI 300 are large-cap blue-chip stocks with lower volatility. In contrast, the CSI 500 primarily consists of mid- and small-cap stocks with a higher proportion of individual investors engaged in emotional trading, leading to greater market volatility and thus stronger profitability. Crude oil, however, as a basic energy commodity, has relatively stable demand, and many market participants engage in hedging transactions, which dampens price fluctuations to some extent, thus resulting in less favorable profitability. Conversely, in the cryptocurrency market where market participants are mainly speculators, prices can fluctuate dramatically with frequent instances of sharp rises and falls, enabling straddle options to demonstrate extremely high profitability.
\begin{figure}[t]

\begin{minipage}[b]{0.32\columnwidth}
\includegraphics[width=\textwidth]{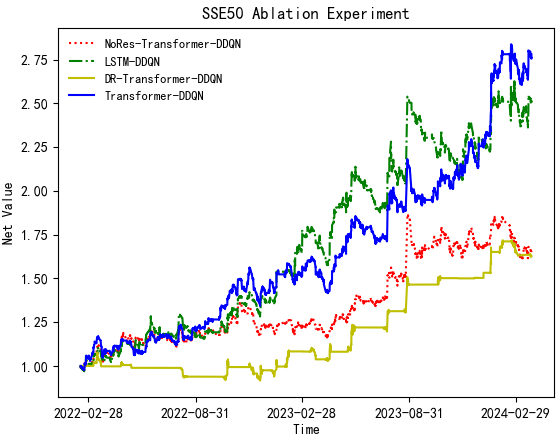}
\end{minipage}
\hfill
\begin{minipage}[b]{0.32\columnwidth}
\includegraphics[width=\textwidth]{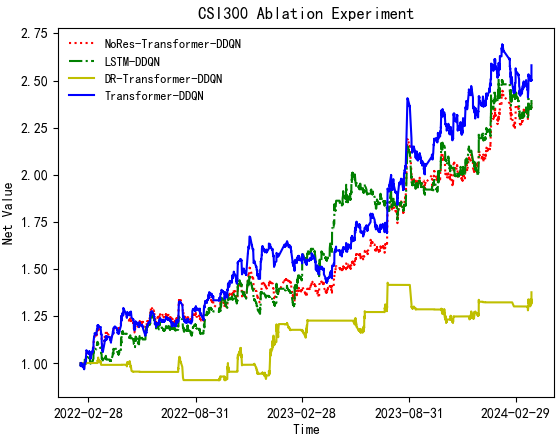}
\end{minipage}
\hfill
\begin{minipage}[b]{0.32\columnwidth}
\includegraphics[width=\textwidth]{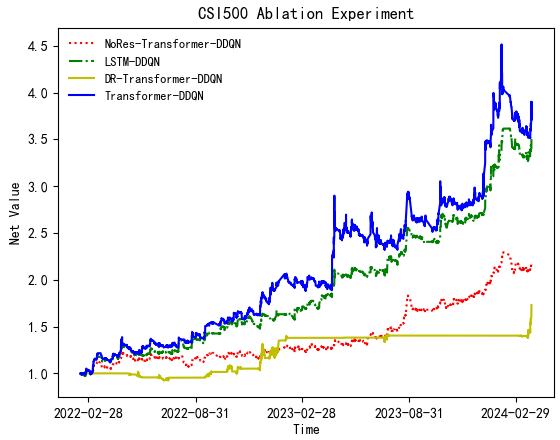}
\end{minipage}

\caption{Ablation result}
\label{AblationResult}
\vspace{-5mm}
\end{figure}
\paragraph{Ablation experiment and analysis}

As shown in Figure \ref{AblationResult} and Table \ref{AblationExperimentsTable}, the performances of the three model variants are inferior to that of the complete Transformer-DDQN model. For NoRes-Transformer-DDQN, removing resistance information causes the model to focus only on current volatility and neglect key regions of recent long-short battles, increasing the model's tendency to misinterpret the volatility. For DR-Transformer-DDQN, since the model constructs straddle option positions, the profits and losses fluctuate during sideways movement, significantly interfering with the model’s learning process. The model fails to capture normal market volatility characteristics effectively, only recognizing sporadic large fluctuations. For LSTM-DDQN, its performance is slightly lower than Transformer-DDQN, with less volatility in its performance curve. By examining trading information, we find that LSTM-DDQN trades more frequently than Transformer-DDQN, resulting in higher transaction fees that erode profits. It suggests that LSTM-DDQN is more sensitive to short-term market fluctuations but overlooks historical volatility information, although it performs better in extremely volatile conditions. This observation is reasoned that Transformer, compared to LSTM, captures long-term dependencies better, effectively filtering out market noise to focus on significant market movements. However, when market volatility decreases, Transformer becomes less responsive with some profit givebacks. The relative strengths and weaknesses of Transformer warrant further research.

\section{Conclusion and future work}
\label{ConclusionAndFuture work}
This paper proposes the Transformer-DoubleDQN model to learn straddle option quantitative trading strategies, focusing on the volatility of trading assets. The model aims to achieve steady returns during normal market fluctuations and excess returns during extreme market movements. Compared to models that trade based on asset price movements, the primary risk source here is the decay of the time value of the option rather than asset price volatility, enabling better risk exposure management. Implied volatility is a crucial factor in option pricing, which typically fluctuates around historical volatility. Due to experimental limitations, this paper assumes implied volatility approximates historical volatility. However, in real markets, implied volatility reflects market sentiment and is difficult to predict. When markets experience significant declines, implied volatility spikes, and option buyers must be aware of the risk of implied volatility dropping in subsequent movements. Integrating implied volatility information into deep reinforcement learning models will be a future research direction.

\bibliographystyle{IEEEtran}
\bibliography{icde25}


\end{document}